\documentclass[12pt]{iopart}

\usepackage{iopams}
\eqnobysec

\usepackage{graphics}
\usepackage{dcolumn}
\usepackage{bm}
\usepackage{hyperref}

\usepackage{amssymb}
\usepackage{amsfonts}

\begin{document}

\title[Potentials in split gate tunneling transistors]{The calculus of the electric potential and field intensity in  multiple electrodes lateral tunneling transistors with double gate}

\author{AS. Spanulescu$^{1}$}

\address{$^{1}$Department of Physics, Hyperion University of Bucharest, Postal code
030629, Bucharest, Romania}


\ead{severspa2004@yahoo.com}
\begin{abstract}
The paper presents an adaptive over-relaxation method for
calculating the electric potential and field intensity, for a
complex tunnel transistor structure involving a split gate and a
shielding boundary. The accuracy and speed of the method has been
numerically tested and found satisfactory for the study of such
devices by calculating the tunneling currents for the obtained
potential distribution.\end{abstract}

\maketitle

\section{Introduction}
\label{} This highly promising tunnel transistor structures have
been intensely investigated in the last decade, for  the future
industrial application. Among the large variety of proposed models
there are some with simple barrier \cite{c1}, and others that
include one or more quantum wells  \cite{c2}. The first type is
based on the modulation of the Fowler-Nordheim tunneling current by
the electric field intensity created by an auxiliary electrode. The
second type is based on the resonance effects between the quantum
wells, as a significant tunneling current appears only if some
quantum states equivalence is satisfied. An auxiliary electrode
controls the resonance conditions, starting or stopping the
tunneling current, so that the structure is easy to be used as a
commutation device. Concerning the geometry, there are also several
types of tunnel transistors. Thus, the most promising seems to be
the vertical transistor where the electrodes are superposed, which
allows a very good control of the thickness of the barrier layer and
hence a good reproducibility for large scale integration
\cite{c1},\cite{c3},\cite{c4}\cite{c7},\cite{c8}. Also, lateral
structures have been proposed, based on a Columbian barrier
generated by a lateral p-n junction, which is easier to control by
the gate electrode \cite{c5},\cite{c6}. There are also some mixed
structures [2] where both vertical and horizontal geometries are
used, and also some special devices as the magnetic tunnel
transistor and the single electron transistor (SET) used as
structures in scientific research \cite{c9}.

The calculus of the tunneling currents in the tunnel transistors
structures implies the previous calculation of the electric
potentials or electric field intensity in all the points of the
active domain, for various potentials applied on the electrodes.

These values will be used as an ansatz for the quantum mechanics
calculations of the tunneling currents using either Schr\"{o}dinger
or Nordheim-Fowler type equations, and hence the static
characteristics of the tunnel transistor for a given geometry and
for certain materials of the structure.

The potentials in a lateral tunnel transistor type structure obeys
the bi-dimensional Laplace equation:
    \begin{equation}
\frac{{\partial ^2 V(x,y)}}{{\partial ^2 x^2 }} + \frac{{\partial ^2
V(x,y)}}{{\partial ^2 y^2 }} = 0
\end{equation}
with the boundary conditions  (Dirichlet type) :
    \begin{equation}
V(x,y)|_{\partial _i D}  = a_i \quad ,i = 1,2,...,n
\end{equation}
where $\partial _i D$
 is one of the $n$
 boundaries of the metallic electrodes or of the whole domain.

For simple systems with a certain symmetry it may be possible to
analytically solve such an equation, but any change in the geometry
would impose a new analytical calculation. For practical purpose,
the mathematical modeling has to be numerically solved so that the
some computer software should be used for a big number of models. In
most of the cases, a rectangular bi-dimensional domain may be
considered and a grid of equidistant points (nodes) chosen. If the
axes origin is a corner of the rectangular domain and if there are
$N$ points in each direction, with the steps:

$h_x  = \frac{{x_{\max } }}{N}$

$h_y  = \frac{{y_{\max } }}{N}$

we may find the values of the potentials in the nodes: $V(x_i ,y_i
)\;;\quad i = 1,2,...,N\;;\quad j = 1,2,...,N$

The finite difference approximation of the eq. (1.1)  leads to:
    \begin{equation}
\begin{array}{l}
 V(x_i ,y_i ) = \frac{{h_y^2 }}{{2(h_x^2  + h_y^2 )}}[V(x_{i - 1} ,y_j ) + V(x_{i + 1} ,y_j )]  \\
 \quad \quad \quad \quad \quad  + \frac{{h_x^2 }}{{2(h_y^2  + h_x^2 )}}[V(x_i ,y_{i - 1} )] + V(x_i ,y_{i + 1} )] \\
 \end{array}
\end{equation}

    If $x_{\max }  = y_{\max } $
 (square domain) we have $h_x  = h_y $
and the potential in a node is the mean of the potentials of its
neighbors.
    \begin{equation}
    \begin{array}{l}
V(x_i ,y_i ) = \frac{1}{4}[V(x_{i - 1} ,y_j ) + V(x_i ,y_{j - 1} )\\
+ V(x_{i + 1} ,y_j ) + V(x_i ,y_{j + 1} )]\quad i,j = 1,2,...,N
  \end{array}
\end{equation}

 The errors are of the order ${\cal O}(h^2 )$
so that an $100 \times 100$ points grid is adequate for most
practical purposes. However, the equations system for such a grid
has $10^4 $ equations and unknowns, which implies increased errors
due their propagation and the often bad conditioning. That is why in
many cases the direct solving of such a system is not the best
choice, and the iterative methods are preferred.

 Thus, a Jacobi iteration \cite{c10} will compute at the step $k + 1$
 the potential value from the neighborhood potentials calculated at step $k$
 :
    \begin{equation}
      \begin{array}{l}
V^{(k + 1)} (x_i ,y_i ) = \frac{1}{4}[V^{(k)} (x_{i - 1} ,y_j ) +
V^{(k)} (x_{i + 1} ,y_j ) \\
+ V^{(k)} (x_i ,y_{j - 1} ) + V^{(k)} (x_i ,y_{j + 1} )]
 \end{array}
\end{equation}

A better convergence is achieved if we include in the $k + 1$ step
the previously calculated values of the same step, according to the
Gauss-Siedel algorithm \cite{c10}:
    \begin{equation}
  \begin{array}{l}
V^{(k + 1)} (x_i ,y_i ) = \frac{1}{4}[V^{(k + 1)} (x_{i - 1} ,y_j )
+ V^{(k + 1)} (x_i ,y_{j - 1} ) \\
+ V^{(k)} (x_{i + 1} ,y_j ) +
V^{(k)} (x_i ,y_{j + 1} )]
 \end{array}
\end{equation}

The convergence may be further increased using an over-relaxation
method, with a relaxation factor $\beta  \in (1,2)$
 that may be experimentally determined:
    \begin{equation}
\begin{array}{l}
 V^{(k + 1)} (x_i ,y_i ) = (1 - \beta )V^{(k)} (x_i ,y_j ) + \beta \frac{1}{4}[V^{(k + 1)} (x_{i - 1} ,y_j ) \\
 + V^{(k + 1)} (x_i ,y_{j - 1} )] + V^{(k)} (x_{i + 1} ,y_j ) + V^{(k)} (x_i ,y_{j + 1} )
 \end{array}
\end{equation}

               This procedure may be much faster than Gauss-Siedel (which is equivalent with the $\beta  = 1$
 case) if a proper  value of $\beta $
 is chosen so that, taking into account the speed of modern computers, it is suitable for tunneling transistors structures.

\section{The electric potential distribution in the lateral tunneling transistor with
four metallic electrodes }

\begin{figure}
\centering
\includegraphics{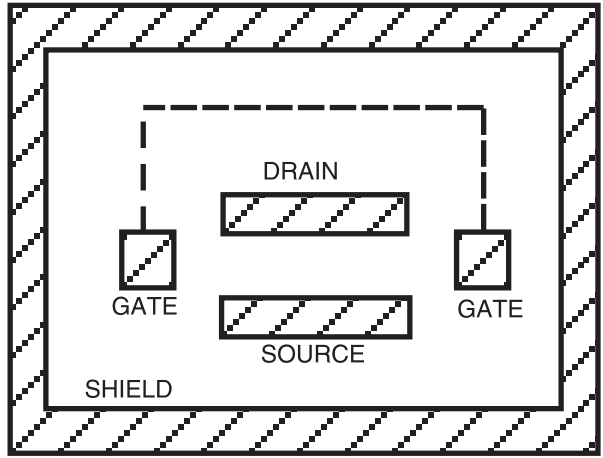}
   \caption{Top view of the proposed lateral transistor structure}
   \label{figura1}
\end{figure}

\begin{figure}
\centering
\includegraphics{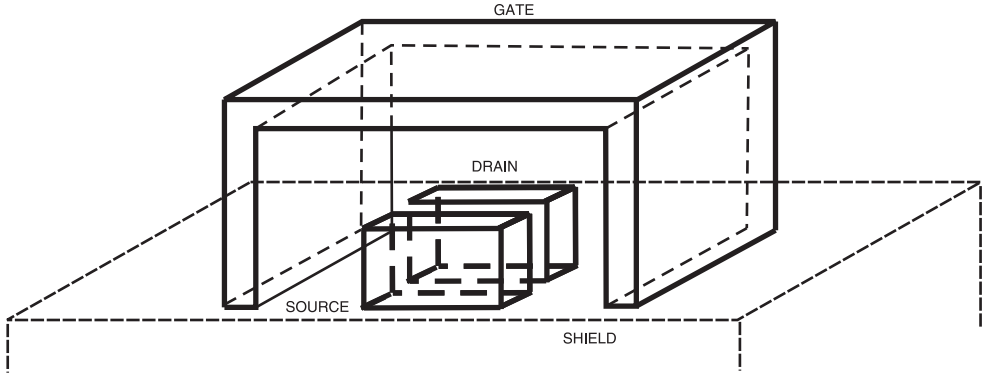}
   \caption{Lateral view of the proposed lateral transistor structure  (the shield is partially shown)}
   \label{figura2}
\end{figure}

We propose  a tunneling transistor structure with coplanar source,
drain and gate as in \fref{figura1}. For a better control efficiency
of the gate, it is split and symmetrically related to the
source-drain region. The two halves are interconnected in a
different plane  at considerable distance from the main plane, to
avoid  tunneling, as in \fref{figura2}.

Also, a grounding metallic shield avoids the influence between
adjacent transistors and establish the boundary conditions for the
second order elliptic partial derivative equation.

This geometry imposes the Dirichlet conditions for the partial
derivative equation. We denote the following boundaries:

$\partial S$    -the source electrode domain and its boundary;

$\partial D$    -the drain electrode domain and its boundary;

$\partial G$    -the gain electrode domain and its boundary;

$\partial Sh$   - the shield electrode domain and its boundary;

Taking into account this geometry, we impose:
    \begin{equation}
V(x_i ,y_j ) = V_S \quad ,\quad \{ x_i ,y_j \}  \in \partial S
\end{equation}

    \begin{equation}
V(x_i ,y_j ) = V_D \quad ,\quad \{ x_i ,y_j \}  \in \partial D
\end{equation}

    \begin{equation}
V(x_i ,y_j ) = V_G \quad ,\quad \{ x_i ,y_j \}  \in \partial G
\end{equation}

    \begin{equation}
V(x_i ,y_j ) = 0\quad ,\quad \{ x_i ,y_j \}  \in \partial Sh
\label{24}
\end{equation}

    If we choose the axes origin in the left lower corner of the domain, the condition \ref{24} becomes:

 $V(x_i ,0) = 0 \quad ,\quad i = 1,2,...,N$

 $V(0,y_j ) = 0 \quad ,\quad j = 1,2,...,N$

 $V(x_i ,N) = 0 \quad ,\quad i = 1,2,...,N$

 $V(N,y_j ) = 0 \quad ,\quad j = 1,2,...,N$

    Considering that the errors are indicated by the values variances from an iteration to another, we have to include in the algorithm the quantity:
    \begin{equation}
\varepsilon _k  = V^{(k + 1)} (x_m ,y_m ) - V^{(k)} (x_m ,y_m )
\end{equation}

The point $\{ x_m ,y_m \} $ will characterize the errors evolution
for the whole domain, as a "mean" value for the potentials
variations. In practice we closed the midpoint of the domain and we
supposed that the errors in all the other points are quite similar.
A choice of a single point for this purpose is necessary for making
a single evaluation at an iteration as a criterium for stopping the
iterative process.

A more precise choice for the point where the error is evaluated
should be one with a maximum absolute value, near one of the
electrodes. Anyway, the errors may be kept much  lower than the
technological dispersion, as our experiments proved.

For accelerating the iterative process, we elaborated an adaptive
algorithm, varying the relaxation factor $\beta $
 in accordance with the variance of $\varepsilon _k $
, taking a medium start value $\beta _0  = 1.5$ .
    \begin{equation}
\beta _k  = \beta _0  + E_{\max } (\varepsilon _{k + 1}  -
\varepsilon _k )
\end{equation}

The typical values of the relaxation factor $\beta $ achieved for
various geometries and electrodes potential is between  1.7  and
1.9. For a grid with $N = 100$, the number of iterations for a $10^{
- 4} $ precision goal is between 200 and 500 and the calculation
time around some minutes.

\section{The electric field intensity  in the lateral tunneling transistor with four metallic electrodes}

If the tunneling current calculus is based on the Schrödinger
equation (the ab-initio method), only the potential distribution in
the structure is needed. However, if the more efficient
Nordheim-Fowler equations are directly used, they need the values of
the electric field intensity.
    It may be calculated from the partial derivatives of the potential:

    \begin{equation}
E_x (x_i ,y_j ) =  - \frac{{\partial V(x,y)}}{{\partial x}}|_{x_i
,y_j }
\end{equation}

    \begin{equation}
E_y  =  - \frac{{\partial V(x,y)}}{{\partial y}}|_{x_i ,y_j }
\end{equation}

    Numerically, the mid-point formula of the derivative may be used, as it provides a good enough precision:
    \begin{equation}
E_x (x_i ,y_j ) =  - \frac{{V(x_{i + 1} ,y_j ) - V(x_{i - 1} ,y_j
)}}{{2h}}
\end{equation}

    \begin{equation}
E_y (x_i ,y_j ) =  - \frac{{V(x_i ,y_{j + 1} ) - V(x_i ,y_{j - 1}
)}}{{2h}}
\end{equation}

    For a better precision we experimented two supplementary methods that we shall describe .
    The first possibility is to use a classical polynomial interpolation (Lagrange-Neville for example):

    \begin{equation}
P_j (x) = \sum\limits_{i = 1}^N {V(x_i ,y_j
)\prod\limits_{\scriptstyle k = 1 \hfill \atop
  \scriptstyle k \ne i \hfill}^N {\frac{{x - x_k }}{{x_i  - x_k }}} } \quad j = 1,2,...,N
\end{equation}

    \begin{equation}
P_i (y) = \sum\limits_{j = 1}^N {V(x_i ,y_j
)\prod\limits_{\scriptstyle k = 1 \hfill \atop
  \scriptstyle k \ne j \hfill}^N {\frac{{y - y_k }}{{y_j  - y_k }};\quad i = 1,2,...,N} }
\end{equation}

    Expanding these expressions in the form:
    \begin{equation}
P_j (x) = \sum\limits_{i = 1}^N {a_{ij} x^i \quad ;\quad j =
1,2,...,N}
\end{equation}

    \begin{equation}
P_i (y) = \sum\limits_{j = 1}^N {b_{ij} y^j \quad ;\quad i =
1,2,...,N}
\end{equation}

the derivatives may be precisely calculated :
    \begin{equation}
P_j^{'} (x) = \sum\limits_{i = 1}^N {ia_{ij} x^{i - 1} } \quad
;\quad j = 1,2,...,N
\end{equation}

    \begin{equation}
P_i^{'} (x) = \sum\limits_{j = 1}^N {jb_{ij} y^{j - 1} \quad ;\quad
i = 1,2,...,N}
\end{equation}

    We obtain the electric field intensity components as
    \begin{equation}
E_x (x_i ,y_j ) = \sum\limits_{i = 1}^N {ia_{ij} x_i^{i - 1} } \quad
ij = 1,2,...,N
\end{equation}

    \begin{equation}
E_y (x_i ,y_j ) = \sum\limits_{j = 1}^N {jb_{ij} x_j^{j - 1} \quad
ij = 1,2,...,N}
\end{equation}

    Although the procedure seems very precise due to the analytical derivation, our tests revealed that it may be used only for low values of $N$
. The problem is generated by the equidistant points grid, which
reveals the Runge phenomenon for high $N$
 values.
    A second solution that was tested was  the choose of the interpolation points according to the Chebyshev (or even Gauss-Lobatto) conditions.
    \begin{equation}
x_k  = {\rm{Int}}\{ N[1 - \cos (2k - 1)\frac{\pi }{{2N}}]\}
\end{equation}

    \begin{equation}
y_l  = {\rm{Int}}\{ N[1 - \cos (2l - 1)\frac{\pi }{{2N}}]\}
\end{equation}

    This eliminates the Runge phenomenon but the complications of the algorithm were not sustained by a significant increase of the precision compared to the mid-point derivative.


\section{Conclusions}

We exemplify the described procedures using a 200x200 grid and the
following potentials:
    $\begin{array}{l}
 V_D  =  - 3V \\
 V_S  = 1V \\
 V_G  = 3V \\
 \end{array}$

 We choose a relaxation factor $\beta  = 1.9$ and the error reference point at {100,100} imposing a 0.01 \%  precision goal. The potential calculated in this point after 330 iterations was -1.42714V and the previous iteration value was -1.42713V. The calculation time on a 3GHz computer in a  Mathematica 7.0 environment was about 310 seconds. Of course, using a C++ environment, the speed would be up to ten times higher.

\begin{figure}
\centering
\includegraphics{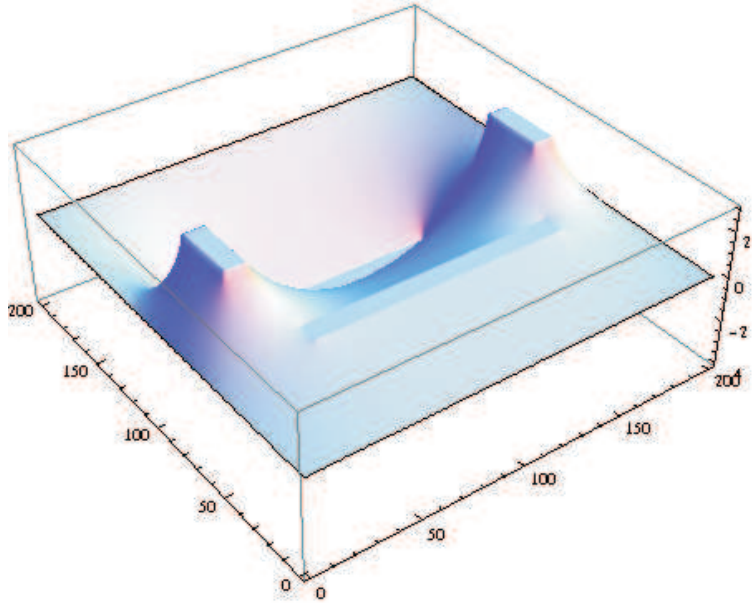}
   \caption{Three-dimensional representation of the potentials - source view}
   \label{figura3}
\end{figure}

\begin{figure}
\centering
\includegraphics{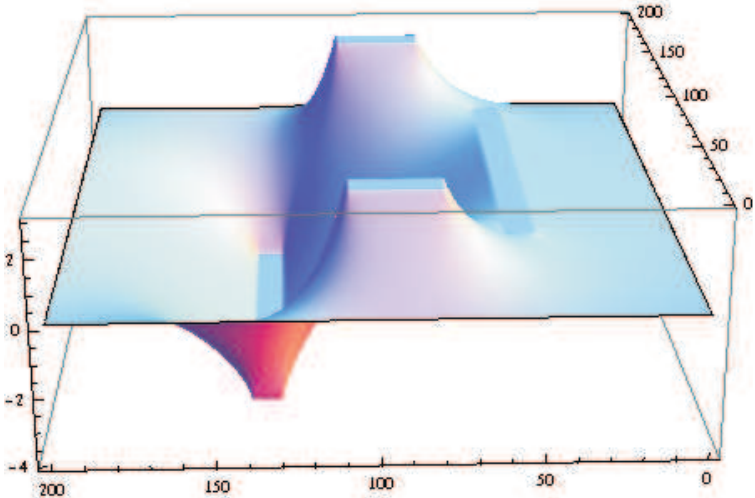}
   \caption{Three-dimensional representation of the potentials - lateral view}
   \label{figura4}
\end{figure}

\begin{figure}
\centering
\includegraphics{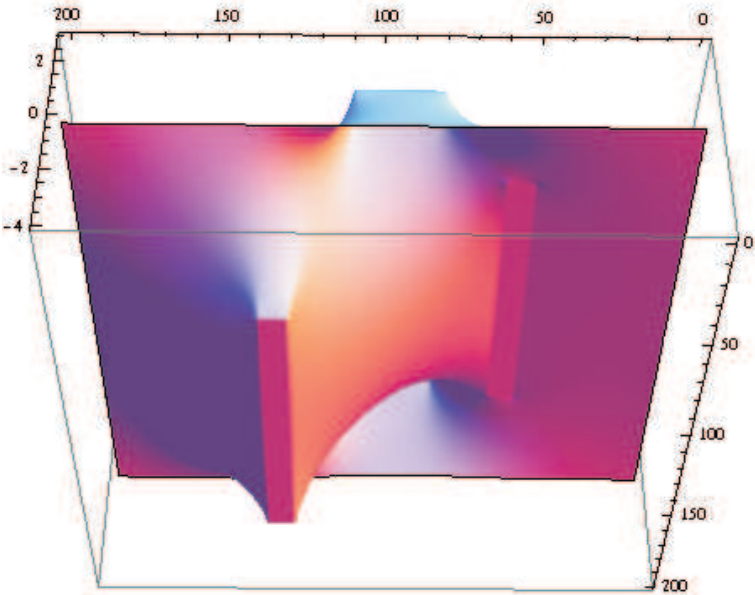}
   \caption{Three-dimensional representation of the potentials - bottom view}
   \label{figura5}
\end{figure}

 A sample of the potential values in a section with x=40 is the following

        V(40,i)={0,0.0244191,0.0487992,0.0731018,0.0972886,0.121322,0.145164,0.168781,0.192136,

0.215195,0.237928,0.260304,0.282294,0.303873,0.325016,0.345701,0.365911,0.385628,

0.40484,0.423538,0.441713,0.459362,0.476485,0.493083,0.509163,0.524733,0.539802,

0.554382,0.568487,0.582129,0.59532,0.60807,0.620383,0.63226,0.643693,0.654667,

        0.665159,0.675138,0.684575,0.693438,0.701706,0.709369,0.71643,0.722907,0.728828,

        0.734227,0.739145,0.743621,0.747696,0.751408,0.754792,0.75788,0.760703,0.763286,

        0.765653,0.767824,0.76982,0.771655,0.773345,0.774903,0.776341,0.77767,0.778898,

        0.780034,0.781086,0.782061,0.782964,0.783802,0.78458,0.785301,0.785971,0.786593,

        0.78717,0.787706,0.788204,0.788665,0.789094,0.789491,0.789859,0.7902,0.790516,

        0.790807,0.791075,0.791323,0.79155,0.791758,0.791948,0.792121,0.792278,0.792419,

        0.792545,0.792657,0.792755,0.792839,0.79291,0.792969,0.793015,0.793049,0.793071,

        0.793081,0.793078,0.793064,0.793038,0.793,0.792949,0.792886,0.79281,0.792721,

        0.792618,0.792501,0.79237,0.792223,0.79206,0.791881,0.791684,0.791469,0.791234,

        0.790979,0.790702,0.790402,0.790077,0.789726,0.789347,0.788938,0.788497,0.788021,

        0.787509,0.786956,0.786362,0.785721,0.785031,0.784287,0.783485,0.78262,0.781687,

        0.78068,0.779593,0.778417,0.777146,0.77577,0.774279,0.772662,0.770906,0.768997,

        0.76692,0.764657,0.762187,0.759488,0.756536,0.753302,0.749756,0.745863,0.741588,

        0.736893,0.731738,0.726088,0.719907,0.713171,0.705863,0.697979,0.689529,0.680533,

        0.671021,0.66102,0.650558,0.639656,0.628326,0.616573,0.604396,0.591788,0.578737,

        0.56523,0.551252,0.53679,0.52183,0.50636,0.490372,0.473858,0.456815,0.439243,

        0.421143,0.402521,0.383385,0.363748,0.343623,0.323027,0.301981,0.280507,0.258628,

        0.236372,0.213767,0.190842,0.167631,0.144164,0.120478,0.0966066,0.0725862,0.0484534

        ,0.0242455,0}

    In figures 3-6 we also show a three-dimensional representation of the calculated potentials with different view points

\begin{figure}
\centering
\includegraphics{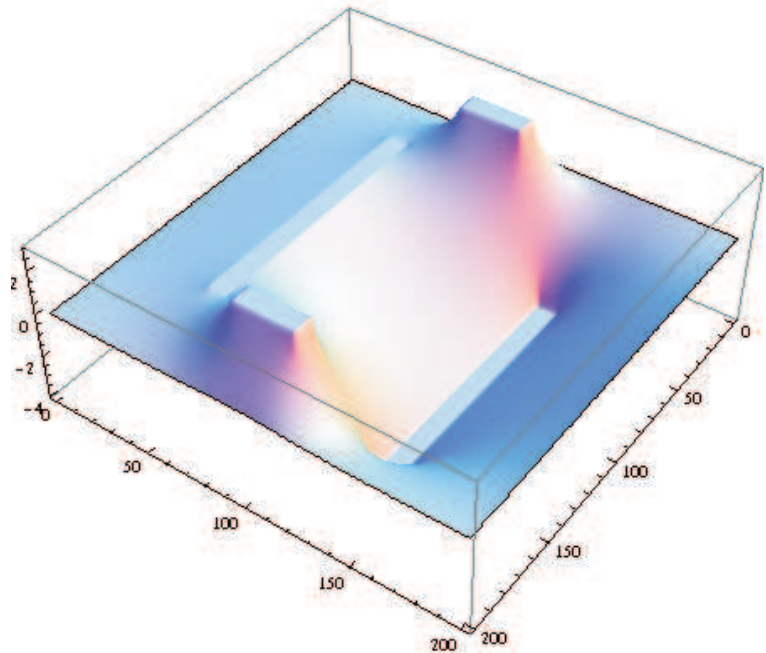}
   \caption{Three-dimensional representation of the potentials - drain view}
   \label{figura6}
\end{figure}

We conclude that the over-relaxation method is well suited for the
calculus of the potential distribution in a multiple electrodes
structure with reasonable speed and accuracy, especially if the
described adaptive algorithm for adjusting the relaxation factor is
used.

\bigskip

\bigskip

\bigskip

\bigskip

\ack This work was supported by The Council of Scientific Research
in Higher Studies in Romania (CNCSIS) under Grant 556/2008.

--------------------------------------------------------------------------------------------

\end{document}